\documentstyle{article}

\author{C.G.Beneventano and E.M.Santangelo\thanks{This work was
partially supported by CONICET and Fundaci\'on Antorchas,
Argentina}\\
Departamento de F\'\i sica - Facultad de Ciencias Exactas - UNLP}
\title{Connection between $\zeta$ and cutoff regularizations of
Casimir energies}

\begin{document}

\maketitle
\begin{abstract}
We study the connection between $\zeta $- and cutoff-regularized
Casimir
energies for scalar fields. We show that, in general, both
regularization
schemes lead to divergent contributions, and to finite parts which do
not
coincide. We determine the relationships among the various
coefficients
appearing in one approach and the other. As an application, we
discuss the
case of scalar fields in $d$-dimensional boxes under periodic
boundary
conditions.
\bigskip

\end{abstract}

\section{Introduction}

The task of extracting physically meaningful results from
ill defined
quantities is a fundamental aspect of quantum field theory. Perhaps
the
simplest example of this situation is the infinite zero point energy
of
quantum fields. In 1948, Casimir showed $\cite{Casimir}$ that
neutral
perfectly conducting parallel plates placed in the vacuum attract
each
other. The basic idea behind the concept of Casimir vacuum energy is
that
quantum fields always exist in the presence of external constraints
and
their zero point field energy is thus modified. Such constraints are
idealized as conditions to be satisfied by modes of the field at the
boundary of a given manifold.

One of the procedures $\cite{Plunien}$ used for computing Casimir
energies
is the direct evaluation of infinite sums over zero modes. These
sums, which
are formally divergent, may be regularized through various techniques
$\cite
{Plunien}$,$\cite{Ambjorn}$.

Recently, some work aiming at understanding the relationship between
the
$\zeta$ function and exponential cutoff  regularizations of Casimir
energies
for scalar fields was performed by the authors of references
$\cite{Svaiter1}
$,$\cite{Svaiter2}$. In particular, in $\cite{Svaiter2}$, the
connection
between both regularizations was established, with the ad-hoc
condition that
the cutoff-regularized energy presents only polar divergencies. The
$\zeta $%
-regularized energy was then shown to be finite and its outcome was
proved
to be identical to the energy regularized via cutoff, with polar
terms
subtracted.

In this paper, we extend the study of the above-mentioned connection
to the
most general case of physical interest : we only restrict the
associated
boundary problem to guarantee that the energy eigenvalues of the
field be
real. Then, we make use of some well known results concerning traces
of heat
kernels and complex powers of differential operators \cite{Gilkey},
\cite
{Seeley}, \cite{Seeley2} for such boundary problems. We apply Mellin
transform techniques to show that both regularizations lead, in
general, to
divergences and that their finite parts do not coincide.

The paper is organized as follows :

Section 2 contains a presentation of the problem to be studied.
There, the
formal expression for the Casimir energy of a scalar field through
mode
summation is given. The associated boundary problem is defined and
$\zeta $
and cutoff regularizations are introduced.

In Section 3 , we present our main result : we show that under
fairly
general conditions on the associated boundary problem
$\cite{Gilkey}$,$\cite
{Seeley}$ (which guarantee the real nature of the energy
eigenvalues), the
regularization via $\zeta $ function does not, in general, lead to a
finite
result and -consequently- the exponential regularization shows not
only
poles but also logarithmic singularities. The relationship among the
various
coefficients in one regularization and the other is established. In
particular, finite parts are seen to differ in a well determined
fashion.

Section 4 contains a simple example of application : the evaluation
of the
Casimir energy for a massive scalar field in a $d$-dimensional box,
subject
to periodic boundary conditions. Both regularization schemes are
applied to
the cases $d=1$ and $d=2$, and their outcome is shown to agree with
our
general result in Section 3. For this particular geometry, and under
periodicity conditions, the exponential and $\zeta $ regularizations
are
seen to be equivalent only after a physically meaningful prescription
is
given in order to eliminate divergencies.

Finally, our conclusions are presented in Section 5.

\section{Casimir energies for scalar fields through mode summation}

The evaluation of Casimir energies through the mode summation method
involves the direct performance of infinite sums over energy
eigenvalues of
the zero point field modes $\cite{Plunien}$, $\cite{Ambjorn}$,
\begin{equation}
\label{E}E_C=\frac 12\sum_n\omega _n\qquad
\end{equation}

The energy eigenvalues, $\omega _n$, depend on the dimension of the
space-time, on the spin of the field under consideration and on the
boundary
condition imposed on it.

Let us consider the case of a free scalar field in a $d+1$-
dimensional
space-time manifold
\begin{equation}
{\cal M}{\em ^{\left( d+1\right) }=R\times M\qquad }
\end{equation}
where $M$ is a smooth compact $d$-dimensional manifold with smooth
boundary $\partial M$.

After separation of variables, the energy eigenvalues turn out to be

\begin{equation}
\omega _n=\lambda _n^{1/2}\qquad
\end{equation}
where the $\lambda _n$ satisfy $\cite{Ambjorn}$ the associated
boundary
problem

\begin{equation}
\label{op}D_B\varphi _n=\left\{ {D\varphi _n=\lambda _n\varphi
_n}\atop{%
BT\varphi _n=0}\right. \qquad
\end{equation}

Here, $D$ is a second order operator on  $M$, $B$ is
a
tangential operator (which we will take to be differential),
defining
boundary conditions and $T$ is the restriction map, which assigns to
each
section its Cauchy data at $\partial M$.

In what follows, we will refer to the boundary problem $\left(
\ref{op}%
\right) $ as $\left( D,B\right) $. It is clear that, in order for the
$%
\omega _n$ to make sense as physical energies, the eigenvalues
$\lambda _n$
must be real and positive, which can be achieved by imposing certain
well
known conditions $\cite{Gilkey}$, $\cite{Seeley}$ (to be specified in
the
next section) on the boundary problem $\left( D,B\right) $. As we
will see,
such conditions also imply that $\omega _n
\begin{array}{c}
\\
\rightarrow  \\
n\rightarrow \infty
\end{array}
\infty $ and that they are $O\left( n^{1/d }\right) $ for $n$
large.

Thus, the mode summation $\left( \ref{E}\right) $ is divergent , and
a
meaning must be given to it through some regularization scheme. In
this
paper we will be concerned with two of these methods : the
exponential
cutoff and $\zeta $ function $\cite{Hawking}$ regularizations.

In the first case, one defines,
\begin{equation}
E_{\exp }\equiv \left. \frac \mu 2\sum\limits_n\frac{\lambda
_n^{1/2}}\mu
e^{-t\frac{\lambda _n^{1/2}}\mu }\right\rfloor _{t=0}=\left. -\frac
\mu
2\frac d{dt}\left( h\left( t,\frac{D_B}{\mu ^2}\right) \right)
\right\rfloor
_{t=0}\qquad
\end{equation}
where

\begin{equation}
\label{h}h\left( t,\frac{D_B}{\mu ^2}\right)
=\sum\limits_ne^{-t\frac{%
\lambda _n^{1/2}}\mu }=Tr\left( e^{-\frac t\mu D_B^{1/2}}\right)
\qquad
\end{equation}
and $\mu $ is a parameter with dimensions of mass, introduced in
order to
render $t$ dimensionless.

As regards $\zeta $ function regularization $\cite{Hawking}$, the
Casimir
energy is defined as
\begin{equation}
\label{z}
\begin{array}{c}
E_\zeta \equiv \left. \frac \mu 2\sum\limits_n\left(
\frac{\lambda _n}{\mu ^2}\right) ^{-\frac s2}\right\rfloor
_{s=-1}=\left.
\frac \mu 2\zeta \left(\frac s2, \frac{D_B}{\mu ^2}\right)
\right\rfloor
_{s=-1} \\  \\
=\left. \frac \mu 2Tr\left( \frac{D_B}{\mu ^2}\right) ^{-\frac
s2}\right\rfloor _{s=-1}
\end{array}
\end{equation}

{}From the already mentioned behaviour of $\lambda _n$ it is easy to
see that
the sum in $\left( \ref{z}\right) $ is convergent for $Re(s)$ large
enough.
So, $\zeta \left( \frac{D_B}{\mu ^2},\frac s2\right) $ is holomorphic
in the
same region. As we will see in the next section, it can be extended
to the
hole $s$ plane as a meromorphic function, with only single poles. The
$\zeta
$ function regularized energy is then defined as the value of this
meromorphic extension at $s=-1$. Here, a parameter $\mu $ has again
been
introduced, this time in order to render the $\zeta $ function
dimensionless
$\cite{Wipf}$.

It will be the subject of the next section to establish the behaviour
of
Casimir energies regularized in both fashions, and to give the
precise
relationship between divergent and finite parts appearing in one
scheme and
the other, thus generalizing the result in reference \cite{Svaiter2}.
We
will also show that both regularizations are not, in general,
equivalent
(i.e., finite parts differ). A particular case will be studied in
Section 4,
where they will be shown to give the same result after an adequate
prescription is imposed to eliminate divergencies.

\section{Equivalence between regularizations}

In this section we present our main result. We study, under fairly
general
conditions (which are those of physical interest), the connection
between
exponential and $\zeta $ function regularizations of Casimir energies
for
scalar fields. We establish the relationships among coefficients
appearing
in one case and the other.

Before going to our main result, we reproduce, without proof, some
well
known facts concerning elliptic boundary problems
$\cite{Gilkey}$,$\cite
{Seeley}$, as applied to the case of interest :

\bigskip\

{\cal Lemma 1}

Let $M$ be a smooth compact $d$-dimensional manifold, with smooth
boundary $%
\partial M$. Let $D$ be an elliptic second order partial
differential
operator, and let $B$ be a tangential differential operator over
$\partial M$%
{}.

If the boundary problem $\left( D,B\right) $ is self-adjoint and
elliptic
with respect to $C-R_{+}$ (i.e., it has an Agmon's cone
$\cite{Agmon}$
including the negative axis), then :

a) We can find a complete orthonormal system $\left\{ \phi
_n\right\}
_{n=1}^\infty $ with $D\phi _n=\lambda _n\phi _n$.

b) $\phi _n$ satisfy the boundary condition $BT\phi _n=0$ (here, $T$
is the
restriction map, which assigns to any smooth section its Cauchy
data).

c) $\lambda _n\in R$ and $\lim _{n\rightarrow \infty }\left| \lambda
_n\right| =\infty $. If we order the $\lambda _n$ such that $\left|
\lambda
_1\right| \leq \left| \lambda _2\right| \leq \cdots $, then there
exists $n_0$
so that $\left| \lambda _n\right| >n^{\frac 2d}$ for $n>n_0$.

d) The $\lambda _n$ are bounded from below and  spec ($D_B$)
is
contained in $\left[ -C,\infty \right] $ for some constant $C$.

(In what follows we will assume, without loosing generality that
spec ($%
D_B$) is positive).

\bigskip\

{\cal Lemma 2}

Under the conditions of the previous lemma $\cite{Seeley}$ :

a) $Y\left( t,\frac{D_B}{\mu
^2}\right) =Tr\left( e^{-t\frac{D_B}{\mu ^2}}\right) $ is
holomorphic in a sector $V_{\theta _0}$ (for some $\theta _0\in
\left(
0,\frac \pi 2\right) $, $V_{\theta _0}=\left\{ t=re^{i\theta }
/
r>0,\left| \theta \right| <\theta _0\right\} $.

b) $Y\left( t,\frac{D_B}{\mu
^2}\right) $ has the asymptotic expansion :$\qquad Y\left(
t\right) \sim \sum\limits_{j=0}^\infty a_jt^{\frac{j-d}2}\quad ,$ for
$%
t\rightarrow 0$ uniformly for $t\in V_\delta $, for each $\delta
<\theta _0$.

Here, the $a_j$ can be evaluated from Seeley's coefficients
$\cite{Seeley}$,
including volume as well as boundary contributions.

c) $Y\left( t,\frac{D_B}{\mu
^2}\right) $ decreases exponentially for $\left| t\right|
\rightarrow \infty $ in $V_\delta $.

\bigskip\

{\cal Lemma 3}

Under the same conditions $\cite{Seeley}$,$\cite{Seeley-Grubb}$ :

a)
\begin{equation}
\Gamma \left( \frac s2\right) \zeta \left( \frac s2,\frac{D_B}{\mu
^2}%
\right) =\Gamma \left( \frac s2\right) Tr\left( \left( \frac{D_B}{\mu
^2}%
\right) ^{-\frac s2}\right) =\int_0^\infty t^{\frac s2-1}Y\left(
t,\frac{D_B}{\mu
^2}\right) dt
\end{equation}
is the Mellin transform of $Y(t,\frac{D_B}{\mu ^2})$. It is
holomorphic for $Re(s)>d$
and
extends to a meromorphic function, with isolated simple poles
:
\begin{equation}
\label{z1}\Gamma \left( \frac s2\right) \zeta \left( \frac
s2,\frac{D_B}{\mu
^2}\right) =\sum\limits_{j=0}^\infty \frac{2a_j}{s+j-d}+r\left(
\frac
s2\right) \qquad
\end{equation}
where $r\left( \frac s2\right) $ is an entire function.

b) For each real $c_1,c_2$ and each $\delta <\theta _0$,
\begin{equation}
\label{propz}\left| \Gamma \left( \frac s2\right) \zeta \left( \frac
s2,
\frac{D_B}{\mu ^2}\right) \right| \leq C\left( c_1,c_2,\delta
\right)
e^{-\delta \left| Im\frac s2\right| }\quad ,\left| Im\frac s2\right|
\geq
1,c_1\leq Re\frac s2\leq c_2
\end{equation}

With these elements at hand, we are now able to prove the following
Lemma,
which is the basis of our main result :

\bigskip\

{\cal Lemma 4}

Under the same assumptions as before :

a) $h\left( t,\frac{D_B}{\mu
^2}\right) =Tr\left( e^{-\frac t\mu D_B^{1/2}}\right)
=\sum\limits_ne^{-t\frac{\lambda _n^{1/2}}\mu }$ has the
asymptotic
expansion
\begin{equation}
\label{exph}
\begin{array}{c}
h\left( t,\frac{D_B}{\mu
^2}\right) \sim \sum\limits_{k=0}^d
\frac{\Gamma \left( \frac{k+1}2\right) }{\Gamma \left( \frac
12\right) }%
a_{d-k}\left( \frac t2\right) ^{-k}+\sum\limits_{k=1}^\infty
\frac{\Gamma
\left( -k+\frac 12\right) }{\Gamma \left( \frac 12\right)
}a_{d+2k}\left(
\frac t2\right) ^{2k} \\  \\
+\sum\limits_{k=1}^\infty
\frac{\left( -1\right) ^{k-1}}{\Gamma \left( \frac 12\right) \Gamma
\left(
k\right) }\left( \frac t2\right) ^{2k-1}\left[ r\left( -k+\frac
12\right)
+a_{d+2k-1}\left( \Psi \left( 1\right) -\sum\limits_{l=1}^{k-2}\frac
1{l-k+1}\right) \right.  \\  \\
\left. +\sum\limits_{j\neq
d+2k-1}\frac{2a_j}{j-d-2k+1}-2a_{d+2k-1}\ln
\left( \frac t2\right) \right]
\end{array}
\end{equation}
for $t\rightarrow 0$, uniformly for $t\in V_\delta $, for each
$\delta
<\theta _0$.

\bigskip\

{\cal Proof}

Notice, in the first place, that
\begin{equation}
\Gamma \left( s\right) \zeta \left( \frac s2,\frac{D_B}{\mu
^2}\right)
=\int_0^\infty t^{s-1}h\left( t,\frac{D_B}{\mu
^2}\right) dt
\end{equation}
is the Mellin transform of $h\left( t,\frac{D_B}{\mu
^2}\right) $. Now,
\begin{equation}
\label{z2}
\begin{array}{c}
\Gamma \left( s\right) \zeta \left( \frac s2,
\frac{D_B}{\mu ^2}\right) =\frac{\Gamma \left( s\right) }{\Gamma
\left(
\frac s2\right) }\left[ \Gamma \left( \frac s2\right) \zeta \left(
\frac s2,
\frac{D_B}{\mu ^2}\right) \right] = \\  \\
=\frac{2^{s-1}}{\sqrt{\pi }}\Gamma \left( \frac{s+1}2\right) \left[
\Gamma
\left( \frac s2\right) \zeta \left( \frac s2,\frac{D_B}{\mu
^2}\right)
\right]
\end{array}
\end{equation}

{}From {\cal Lemma 3 }a), and the well known singularity structure of
$\Gamma
\left( \frac{s+1}2\right) $, it turns out that $\left(
\ref{z2}\right) $ is
holomorphic for $Res>d$, and
\begin{equation}
\label{h1}h\left( t,\frac{D_B}{\mu
^2}\right) =\frac 1{2\pi i}\int\limits_{c-i\infty
}^{c+i\infty }ds\quad t^{-s}\frac{2^{s-1}}{\sqrt{\pi }}\Gamma \left(
\frac{%
s+1}2\right) \left[ \Gamma \left( \frac s2\right) \zeta \left( \frac
s2,
\frac{D_B}{\mu ^2}\right) \right] \quad ,\quad c>d
\end{equation}

Moreover, from {\cal Lemma 3} b), together with the fact that $\Gamma
\left(
\frac{s+1}2\right) $ is $O\left( e^{\left( -\frac \pi 2+\epsilon
\right)
\left| Im\frac s2\right| }\right) $, for any $\epsilon >0$, an
asymptotic
expansion for $h\left( t,\frac{D_B}{\mu ^2}\right) $ can be obtained
by shifting the
contour of
integration in $\left( \ref{h1}\right) $ past the poles of $\Gamma
\left(
\frac{s+1}2\right) \left[ \Gamma \left( \frac s2\right) \zeta \left(
\frac
s2,\frac{D_B}{\mu ^2}\right) \right] $. These poles are located at
$s=d-j$.

For $s=d-j=k\geq 0\quad \left( j\leq d\right) $ they are simple
poles, and
they contribute to the Cauchy integral with
\begin{equation}
\label{res1}\frac{\Gamma \left( \frac{k+1}2\right) }{\Gamma \left(
\frac
12\right) }a_{d-k}\left( \frac t2\right) ^{-k}\qquad ,\ k=0,1,\ldots
,d
\end{equation}

For $s=d-j=-2k$ \quad $\left( k=1,2,\ldots \right) $ they are also
simple,
and their contribution is
\begin{equation}
\label{res2}\frac{\Gamma \left( -k+\frac 12\right) }{\Gamma \left(
\frac
12\right) }a_{d+2k}\left( \frac t2\right) ^{2k}\qquad ,k=1,2,\ldots
\end{equation}

For $s=d-j=-\left( 2k-1\right) \quad \left( k=1,2,\ldots \right) $
they are
simple and double poles, which contribute :
\begin{equation}
\label{res3}\frac{\left( -1\right) ^{k-1}}{\Gamma \left( \frac
12\right)
\Gamma \left( k\right) }\left( \frac t2\right) ^{2k-1}\left[ r\left(
-k+\frac 12\right) +\sum\limits_{j\neq
d+2k-1}\frac{2a_j}{j-d-2k+1}\right]
\end{equation}
\begin{equation}
\label{res4}\frac{\left( -1\right) ^k}{\Gamma \left( \frac 12\right)
\Gamma
\left( k\right) }\left( \frac t2\right) ^{2k-1}a_{d+2k-1}\left[ 2\ln
\left(
\frac t2\right) -\left( \Psi \left( 1\right)
-\sum\limits_{l=1}^{k-2}\frac
1{l-k+1}\right) \right]
\end{equation}
(Notice that the last sum in $\left( \ref{res4}\right) $ is to be
included
whenever it makes sense).

So, shifting the path of integration in $\left( \ref{h1}\right) $ up
to, and
including, the singularity at $s=-2K$ we have,
\begin{equation}
\begin{array}{c}
h\left( t,\frac{D_B}{\mu ^2}\right) =\sum\limits_{k=0}^d
\frac{\Gamma \left( \frac{k+1}2\right) }{\Gamma \left( \frac
12\right) }%
a_{d-k}\left( \frac t2\right) ^{-k}+\sum\limits_{k=1}^K\frac{\Gamma
\left(
-k+\frac 12\right) }{\Gamma \left( \frac 12\right) }a_{d+2k}\left(
\frac
t2\right) ^{2k} \\  \\
+\sum\limits_{k=1}^K\frac {\left( -1\right)
^{k-1}}{\Gamma \left( \frac 12\right) \Gamma
\left(
k\right) }\left( \frac t2\right) ^{2k-1}\left[
r\left( -k+\frac 12\right) +a_{d+2k-1}\left( \Psi \left( 1\right)
-\sum\limits_{l=1}^{k-2}\frac 1{l-k+1}\right) \right. \\
\\
\left. +\sum\limits_{j\neq
d+2k-1}\frac{2a_j}{j-d-2k+1%
}-2a_{d+2k-1}\ln \left( \frac t2\right) \right]
+\rho
_K\left( t\right)
\end{array}
\end{equation}

The remainder $\rho _K\left( t\right) $ is given by an integral like
$\left(
\ref{h1}\right) $, but with $c<-2K$ which, as a result of {\cal Lemma
3} b)
and the estimate for $\left| \Gamma \left( \frac{s+1}2\right) \right|
$
already discussed, is $O\left( \left| \frac t2\right|
^{2K+1-\epsilon
}\right) $, which completes the proof.
\bigskip

This asymptotic development can be differentiated term by term, to
obtain an
asymptotic development for $\frac{dh}{dt}$. When evaluated at $t=0$,
it
gives for the Casimir energy,
\begin{equation}
\label{Eexp}
\begin{array}{c}
E_{\exp }=\left. -\frac \mu 2
\frac{dh\left( t,\frac{D_B}{\mu ^2}\right) }{dt}\right\rfloor
_{t=0}=-\frac \mu 2\left.
\sum\limits_{k=1}^d\left( -k\right) \frac{\Gamma \left(
\frac{k+1}2\right) }{%
2^{-k}\Gamma \left( \frac 12\right) }a_{d-k}\left( t\right)
^{-k-1}\right\rfloor _{t=0} \\  \\
-\frac \mu {4\Gamma \left( \frac 12\right) }\left[ r\left( -\frac
12\right)
+a_{d+1}\left( \Psi \left( 1\right) -2\right) +2\sum\limits_{j\neq
d+1}\frac{%
a_j}{j-d-1}\right] +\frac \mu 2\frac{a_{d+1}}{\Gamma \left( \frac
12\right) }%
\ln \left( \frac t2\right)
\end{array}
\end{equation}

As concerns the Casimir energy regularized via zeta function, from
$\left(
\ref{z1}\right) $ it can be seen to be given by
\begin{equation}
\label{Ez}
\begin{array}{c}
E_\zeta =\left. \frac \mu 2\zeta \left( \frac s2,
\frac{D_B}{\mu ^2}\right) \right\rfloor _{s=-1}=\frac \mu {2\Gamma
\left(
-\frac 12\right) }\sum\limits_{j\neq d+1}\frac{2a_j}{j-d-1}+\frac
\mu
{2\Gamma \left( -\frac 12\right) }r\left( -\frac 12\right) \\  \\
+\left. \frac \mu {\Gamma \left( \frac s2\right) }
\frac{a_{d+1}}{s+1}\right\rfloor _{s=-1}= \\  \\
=-\frac \mu {2\Gamma \left( \frac 12\right) }\sum\limits_{j\neq d+1}

\frac{a_j}{j-d-1}-\frac \mu {4\Gamma \left( \frac 12\right) }r\left(
-\frac
12\right) +\frac \mu {2\Gamma \left( \frac 12\right) }a_{d+1}\left(
\frac{%
\Psi \left( 1\right) }2+1-\ln 2\right) \\  \\
-\left. \frac \mu {2\Gamma \left( \frac 12\right)
}\frac{a_{d+1}}{s+1}%
\right\rfloor _{s=-1}
\end{array}
\end{equation}

{}From $\left( \ref{Eexp}\right) $ and $\left( \ref{Ez}\right) $ the
following
conclusions concerning Casimir energies for scalar fields in a
$d+1$%
-dimensional space-time can be drawn :

1) Both regularization methods do, in principle, give rise to
divergent
contributions. If the coefficient $a_{d+1}$ vanishes, the zeta
function
regularization gives a finite result, which coincides with the finite
part
of the energy obtained through exponential regularization. On the
other
hand, this last regularization method presents poles of order
2, 3, $\ldots $, $%
d+1$, the coefficient of the pole of order $k+1$ being $\Gamma
\left(
k+1\right) $ times the residue of $\frac \mu 2\zeta \left( \frac
s2,\frac{D_B}{\mu ^2}%
\right) $ at $s=k$ $\left( k=1,\ldots ,d\right) $.

2) In the general case $\left( a_{d+1}\neq 0\right) $, the
exponential
regularization shows - appart from polar singularities - a
logarithmic
divergence , with a coefficient that equals minus the residue of
$\frac \mu 2\zeta
\left( \frac s2,\frac{D_B}{\mu ^2}\right) $ at $s=-1$. Moreover, the
finite
parts appearing in one and the other regularization scheme then
differ by
terms proportional to $a_{d+1}$. The difference between the finite
part
obtained through exponential regularization and the one obtained via
$\zeta $
is given by
\begin{equation}
\label{dif}-\frac \mu 2\frac{a_{d+1}}{\sqrt{\pi }}\Psi \left(
1\right)
=\frac \mu 2\frac{a_{d+1}}{\sqrt{\pi }}\gamma
\end{equation}
where $\gamma $ is the Euler-Mascheroni constant.

Before ending this section, it is worth pointing out that our results
are,
of course, valid in the case of a boundaryless manifold $M$. In this
case,
the conditions on the boundary problem reduce to the requirement that
the
operator $D$ be self adjoint, with a positive definite principal
symbol. The
coefficients $a_j$ then include only volume contributions, which
vanish for $%
j$ odd $\cite{Gilkey}$, $\cite{Seeley2}$.

\section{A simple example: massive scalar field in a $d$-dimensional
box}

As a simple example of the results just obtained we study, in this
section,
the Casimir energy of a massive scalar field through $\zeta $ and
exponential regularizations, and compare the results with the
general
predictions just made. We will consider the field to satisfy the
Klein-Gordon equation
\begin{equation}
\left( \partial ^2+m^2\right) \varphi \left( x\right) =0
\end{equation}
inside a $d$-dimensional spatial box of finite dimensions
$L_1,L_2,\ldots
,L_d\quad \left( d\geq 1\right) $. Moreover, periodic boundary
conditions
\begin{equation}
\varphi \left( t,L_i\right) =\varphi \left( t,0\right) \qquad ,\
i=1,\ldots
,d
\end{equation}
will be imposed in each spatial direction. (Notice that this problem
is
equivalent to a boundaryless one).

After separation of variables, the modes of the field are easily seen
to be
given by the square roots of the eigenvalues of the $d$-dimensional
Laplacian ($D_{per}$).
\begin{equation}
\omega _{n_1\ldots n_d}=\left[ m^2+\left( \frac{2n_1\pi
}{L_1}\right)
^2+\ldots +\left( \frac{2n_d\pi }{L_d}\right) ^2\right] ^{1/2}\qquad
,\
n_1,\ldots n_d\in Z
\end{equation}
(In the massless case, the mode $n_1=\ldots =n_d=0$ must be excluded,
since
it gives no contribution to the Casimir energy).

A meromorphic extension for $\Gamma \left( \frac s2\right) \zeta
\left(
\frac s2,\frac{D_{per}}{\mu ^2}\right) $ can be obtained through
Jacobi's
inversion formula. Such extension is given by $\cite{Ambjorn}$%
\begin{equation}
\label{zper}
\begin{array}{c}
\Gamma \left( \frac s2\right) \zeta \left( \frac s2,
\frac{D_{Per}}{\mu ^2}\right) =\frac{L_1\ldots L_d}{\pi ^{\frac
d2}}\left(
\frac \mu 2\right) ^d\left[ \left( \frac m\mu \right) ^{-\left(
s-d\right)
}\Gamma \left( \frac{s-d}2\right) \right.  \\  \\
\left. +2\left( \frac \mu m\right) ^{
\frac{s-d}2}\sum\limits_{n_1=-\infty }^\infty \cdots
\sum\limits_{n_d=-\infty }^\infty \ ^{\prime }\left[ \left(
\frac{n_1L_1\mu }%
2\right) ^2+\ldots +\left( \frac{n_dL_d\mu }2\right) ^2\right]
^{\frac{s-d}%
4}\right.  \\  \\
\left. K_{^{\frac{d-s}2}}\left( 2m\left( \left(
\frac{n_1L_1}2\right)
^2+\ldots +\left( \frac{n_dL_d}2\right) ^2\right) ^{\frac 12}\right)
\right]
\end{array}
\end{equation}
where the prime indicates that the term where all $n_i=0$ is to be
omitted.

The last term in $\left( \ref{zper}\right) $ is analytic in the hole
$s$
plane. The first one has poles at $s=d-2k\quad \left( k=0,1,\ldots
\right) $.

Thus, comparison with $\left( \ref{z1}\right) $ shows that, in this
case, $%
a_j=0$ for $j$ odd, which is consistent with our comment at the end
of the
previous section. As regards $a_{2k}$, they can be easily seen to be
given
by
\begin{equation}
\label{aj}a_{j=2k}=\frac{\left( -1\right) ^k}{k!}\left( \frac \mu
2\right)
^d \frac{L_1\ldots L_d}{\pi ^{\frac d2}}\left( \frac m\mu \right)
^{2k}\qquad ,\ k=0,1,\ldots
\end{equation}

At this stage, some general conclusions can be drawn (for $m\neq 0$),
from
our result in Section 3 :

- If the space is even dimensional, then $a_{d+1}=0$; the Casimir
energy
will be finite when calculated through $\zeta $ function
regularization,
while the exponential regularization will show poles. Finite parts
will then
coincide.

- On the other hand, if $d$ is odd, $a_{d+1}\neq 0$ . Both
regularizations
will in this case present divergencies. These will show up as a pole
at $s=-1
$ in the $\zeta $-regularized version, and a logarithmic singularity
as well
as poles in the exponential one. Finite parts will differ by $\left(
\ref
{dif}\right) $. However, being the divergent terms proportional to
the
volume of the box, they can be subtracted through a physically
meaningful
prescription ($E
\begin{array}{c}
\\
\rightarrow  \\
L_1\ldots L_d\rightarrow \infty
\end{array}
0$) $\cite{Casimir}$,$\cite{Ambjorn}\cite{Eliz-Rom}$. This same
prescription
leaves finite results which are coincident; it is in this sense that
equivalence between both regularizations is to be understood in this
particular example.

It should be remarked here that, being all $a_j$ $\left( j\neq
0\right) $
proportional to positive powers of the mass, the massless case is
particular
: in such case, the $\zeta $ function will only present a pole at
$s=d$, and
the $\zeta $-regularized Casimir energy will thus be finite in any
dimension, while the exponential regularization will only show a pole
of
order $d+1$, both finite parts being coincident.

\subsection{Casimir energy for $d=1$}

{}From $\left( \ref{Ez}\right) $ and $\left( \ref{zper}\right) $, the
$\zeta $%
-regularized Casimir energy turns out to be
\begin{equation}
E_\zeta ^{\left( 1\right) }=-\frac m\pi \sum\limits_{n=1}^\infty
\frac
1nK_1\left( nmL\right) +\left. \frac{L\mu ^2}{4\sqrt{\pi }}\left(
\frac m\mu
\right) ^{1-s}\frac{\Gamma \left( \frac{s-1}2\right) }{\Gamma \left(
\frac
s2\right) }\right\rfloor _{s=-1}
\end{equation}
which is divergent at $s=-1$ as already discussed.

By developing the last term around $s=-1$, we get
\begin{equation}
\label{Ez1}E_\zeta ^{\left( 1\right) }=-\frac m\pi
\sum\limits_{n=1}^\infty
\frac 1nK_1\left( nmL\right) +\frac{m^2L}{4\pi }\left( \left. \frac
1{s+1}\right\rfloor _{s=-1}-\ln \left( \frac m{2\mu }\right) -\frac
12\right)
\end{equation}

As regards exponential regularization
\begin{equation}
E_{\exp }^{\left( 1\right) }=\left. -\frac \mu 2\frac d{dt}\left(
\sum_{n=-\infty }^\infty e^{-t\left( \left( \frac{2n\pi }{L\mu
}\right)
^2+\left( \frac m\mu \right) ^2\right) ^{\frac 12}}\right)
\right\rfloor
_{t=0}
\end{equation}
The series can be evaluated making use of Poisson's summation formula
(see
Appendix 1) and, after differentiating, one gets
\begin{equation}
\label{Eexp1}
\begin{array}{c}
E_{\exp }^{\left( 1\right) }=-\frac m\pi \sum\limits_{n=1}^\infty
\frac
1nK_1\left( nmL\right) \\
\\
\left. +\frac{m^2L}{4\pi }\left( -\ln \left( t\right) -\ln \left(
\frac
m{2\mu }\right) +2\left( \frac{mt}\mu \right) ^{-2}-\gamma -\frac
12\right)
\right\rfloor _{t=0}
\end{array}
\end{equation}

Comparison among the various coefficients (both in divergent as
well as
in finite parts) in $\left( \ref{Ez1}\right) $ and $\left(
\ref{Eexp1}%
\right) $ shows a complete agreement with our results in Section 3.
When the
prescription $E^{\left( 1\right) }
\begin{array}{c}
\\
\rightarrow  \\
L\rightarrow \infty
\end{array}
0$ is
imposed, the physically meaningful Casimir energy turns out to be
\begin{equation}
E_{Cas}^{\left( 1\right) }=-\frac m\pi \sum\limits_{n=1}^\infty
\frac
1nK_1\left( nmL\right)
\end{equation}
in the framework of both regularization schemes. This remaining
finite
result can be easily seen to decay exponentially with $L$. On the
other hand,
divergences as well as finite parts proportional to $L$ have been
subtracted
as a consequence of the forementioned prescription, which amounts to
adding
a "constant " to the energy density.

\subsection{Casimir energy for $d=2$}

Again, from $\left( \ref{Ez}\right) $ and $\left( \ref{zper}\right)
$, the $%
\zeta $-regularized Casimir energy is given by
\begin{equation}
\label{Ez2}
\begin{array}{c}
E_\zeta ^{\left( 2\right) }=-
\frac{L_1L_2}{2^3\pi ^{\frac 32}}\sum\limits_{n_1=-\infty }^\infty
\sum\limits_{n_2=-\infty }^\infty \ ^{\prime }\left( \sqrt{\left(
\frac{%
n_1L_1}2\right) ^2+\left( \frac{n_2L_2}2\right) ^2}\right) ^{-\frac
32} \\
\\
K_{\frac 32}\left( 2m
\sqrt{\left( \frac{n_1L_1}2\right) ^2+\left( \frac{n_2L_2}2\right)
^2}%
\right)  \\  \\
+
\frac{\mu ^3}{2^3}\frac{L_1L_2}\pi \left. \left( \frac m\mu \right)
^{-\left( s-2\right) }\frac{\Gamma \left( \frac{s-2}2\right) }{\Gamma
\left(
\frac s2\right) }\right\rfloor _{s=-1}= \\  \\
=-
\frac{L_1L_2}{2^3\pi ^{\frac 32}}\sum\limits_{n_1=-\infty }^\infty
\sum\limits_{n_2=-\infty }^\infty \ ^{\prime }\left( \sqrt{\left(
\frac{%
n_1L_1}2\right) ^2+\left( \frac{n_2L_2}2\right) ^2}\right) ^{-\frac
32}K_{\frac 32}\left( 2m\sqrt{\left( \frac{n_1L_1}2\right) ^2+\left(
\frac{%
n_2L_2}2\right) ^2}\right)  \\  \\
-\frac{L_1L_2m^3}{12\pi }
\end{array}
\end{equation}

The exponentially regularized Casimir energy is
\begin{equation}
E_{\exp }^{\left( 2\right) }=\left. -\frac \mu 2\frac d{dt}\left(
\sum_{n_1=-\infty }^\infty \sum_{n_2=-\infty }^\infty e^{-t\left(
\left(
\frac{2n_1\pi }{L_1\mu }\right) ^2+\left( \frac{2n_2\pi }{L_2\mu
}\right)
^2+\left( \frac m\mu \right) ^2\right) ^{\frac 12}}\right)
\right\rfloor
_{t=0}
\end{equation}
The double series can again be calculated by repeated use of
Poisson's
formula (see Appendix 2). After differentiating, we get
\begin{equation}
\label{Eexp2}
\begin{array}{c}
E_{\exp }^{\left( 2\right) }=-
\frac{L_1L_2}{2^3\pi ^{\frac 32}}\sum\limits_{n_1=-\infty }^\infty
\sum\limits_{n_2=-\infty }^\infty \ ^{\prime }\left( \sqrt{\left(
\frac{%
n_1L_1}2\right) ^2+\left( \frac{n_2L_2}2\right) ^2}\right) ^{-\frac
32} \\
\\
K_{\frac 32}\left( 2m
\sqrt{\left( \frac{n_1L_1}2\right) ^2+\left( \frac{n_2L_2}2\right)
^2}%
\right)  \\  \\
-\frac{L_1L_2m^3}{12\pi }+\left. \frac{L_1L_2\mu ^3}{2\pi
t^3}\right\rfloor
_{t=0}
\end{array}
\end{equation}

As predicted, the $\zeta $ regularization gives a finite result,
which
coincides with the finite part in the exponential regularization.
This last
regularization presents a single pole of order $d+1=3$, whose
coefficient
coincides with $\Gamma \left( 3\right) $ times the residue of
$E_\zeta
^{\left( 2\right) }$ at $s=2$ as expected. After applying the
prescription $E^{\left( 2\right)}
\begin{array}{c}
\\
\rightarrow  \\
L_1 L_2\rightarrow \infty
\end{array}
0$ the
divergence is eliminated, and a finite piece -proportional to the
volume- is
also discarded; we thus get
\begin{equation}
\begin{array}{c}
E_{Cas}^{\left( 2\right) }=-
\frac{L_1L_2}{2^3\pi ^{\frac 32}}\sum\limits_{n_1=-\infty }^\infty
\sum\limits_{n_2=-\infty }^\infty \ ^{\prime }\left( \sqrt{\left(
\frac{%
n_1L_1}2\right) ^2+\left( \frac{n_2L_2}2\right) ^2}\right) ^{-\frac
32} \\
\\
K_{\frac 32}\left( 2m\sqrt{\left( \frac{n_1L_1}2\right) ^2+\left(
\frac{%
n_2L_2}2\right) ^2}\right)
\end{array}
\end{equation}
which, as in the $d=1$ case, decays exponentially when the volume
increases.

\section{Conclusions}

In this paper, we have studied the connection between $\zeta $ and
cutoff
regularizations of Casimir energies for scalar fields in a space-time
$%
R\times M$, with $M$ a $d$-dimensional manifold with or without
boundary.

Under fairly general conditions on the associated boundary problem
(which
are those of physical interest), we have shown that, in general,
both
regularizations lead to divergent terms. These divergencies appear as
a
simple pole when regularizing via $\zeta $, and are logarithmic as
well as
polar in the exponential regularization. Moreover, finite parts do
not in
general coincide. We have also determined the precise relationship
among the
various coefficients appearing in one case and the other.

As an example of application, we have evaluated Casimir energies for
a
scalar field in a $d$-dimensional box, under periodic boundary
conditions.
In this particular example, the $\zeta $ function turns out to be
finite, in
the massive case, for $d$ even and, in the massless case, for any
dimension.
For whatever dimension, both regularizations have been shown to be
equivalent
once the same prescription ($E_C
\begin{array}{c}
\\
\rightarrow  \\
V\rightarrow \infty
\end{array}
0$, with $%
V$ the volume of the box) is imposed to eliminate infinities. We
have
performed the calculation of the energy with exponential
regularization, and
we have verified the agreement with our general result in the cases
$d=1$
and $2$. Although we haven't found the energy via Poisson's formula
for $d>2$
(the process becomes increasingly tedious as the space dimension
grows up),
it is possible, by using the relationships among  coefficients
determined
in Section 3, to obtain the exact result of the exponential
regularization
for whatever dimension, from the energy obtained via
zeta function.

The extension of these results to fields with other spins is
at
present under study.

\bigskip\

The authors acknowledge M. De Francia, H. Falomir, R.E. Gamboa
Sarav\'{\i}  and M.A. Muschietti for useful comments and discussions.

\section*{Appendix 1- Poisson sum for $d=1$}

In this appendix, we derive $\left( \ref{Eexp1}\right) $ for the
exponentially regularized Casimir energy, making use of Poisson's
formula :
\begin{equation}
\label{sum}\sum_{n=-\infty }^\infty f\left( n\right)
=\sum_{p=-\infty
}^\infty c_p
\end{equation}
with
\begin{equation}
\label{cp}c_p=\int_{-\infty }^\infty dxe^{2\pi ipx}f\left( x\right)
\end{equation}

When applied to the calculation of $h\left( t,\frac{D_B}{\mu
^2}\right) $,
it gives
\begin{equation}
\label{sum1}h\left( t,\frac{D_B}{\mu ^2}\right) =\sum_{n=-\infty
}^\infty
e^{-t\left( \left( \frac{2n\pi }{L\mu }\right) ^2+\left( \frac m\mu
\right)
^2\right) ^{\frac 12}}=\sum_{p=-\infty }^\infty c_p\left( t\right)
\end{equation}
where
\begin{equation}
\label{cp1}
\begin{array}{c}
c_p\left( t\right) =\int_{-\infty }^\infty dxe^{2\pi ipx-t\left(
\left(
\frac{2n\pi }{L\mu }\right) ^2+\left( \frac m\mu \right) ^2\right)
^{\frac
12}}= \\  \\
=
\frac{L\mu }\pi \int_0^\infty dx\cos \left( L\mu px\right)
e^{-t\left(
x^2+\left( \frac m\mu \right) ^2\right) ^{\frac 12}}= \\  \\
=\frac{mL}{\pi \sqrt{1+\left( \frac{L\mu p}t\right) ^2}}K_1\left(
\frac m\mu
\sqrt{t^2+\left( L\mu p\right) ^2}\right)
\end{array}
\end{equation}

By replacing $\left( \ref{cp1}\right) $ into $\left(
\ref{sum1}\right) $ and
taking the derivative, we get
\begin{equation}
-\frac \mu 2\frac d{dt}h\left( t,\frac{D_B}{\mu ^2}\right) =-\frac
\mu
2\frac d{dt}\left( 2\sum_{p=1}^\infty c_p\left( t\right) +c_0\left(
t\right)
\right)
\end{equation}

Recurrence relations and ascending series for modified Bessel
functions then
give $\left( \ref{Eexp1}\right) $.

\section*{Appendix 2- Poisson sums for $d=2$}

In order to derive $\left( \ref{Eexp2}\right) $, we make repeated use
of
Poisson's formula, as given by $\left( \ref{sum}\right) $ and $\left(
\ref
{cp}\right) $.

In this case we have
\begin{equation}
h\left( t,\frac{D_B}{\mu ^2}\right) =\sum_{n_1=-\infty }^\infty
\sum_{n_2=-\infty }^\infty e^{-t\left( \left( \frac{2n_1\pi }{L_1\mu
}%
\right) ^2+\left( \frac{2n_2\pi }{L_2\mu }\right) ^2+\left( \frac
m\mu
\right) ^2\right) ^{\frac 12}}
\end{equation}

We first perform the sum over $n_2$ in the same fashion as in
Appendix 1, to
obtain
\begin{equation}
\begin{array}{c}
h\left( t,\frac{D_B}{\mu ^2}\right) =\\ \\
\sum_{n_1=-\infty }^\infty
\sum_{p=-\infty }^\infty \frac{L_2\mu t }\pi \frac{\sqrt{\left(
\frac{2n_1\pi
}{L_1\mu }\right) ^2+\left( \frac m\mu \right) ^2}}{\sqrt{t^2+\left(
L_2\mu
p\right) ^2}}K_1\left( \sqrt{\left[\left( \frac{2n_1\pi }{L_1\mu
}\right)
^2+\left( \frac m\mu \right) ^2\right]\left[t^2+\left( L_2\mu
p\right)
^2\right]}\right)
\end{array}
\end{equation}

Now, due to the convergence properties of the double sum, the
summation
order can be interchanged, and Poisson's formula can again be used to
obtain
\begin{equation}
\label{sum2}h\left( t,\frac{D_B}{\mu ^2}\right) =\sum_{p=-\infty
}^\infty
\frac{L_2\mu }\pi \frac t{\sqrt{t^2+\left( L_2\mu p\right) ^2}%
}\sum_{k=-\infty }^\infty c_k\left( t\right)
\end{equation}
with
\begin{equation}
\begin{array}{c}
c_k\left( t\right) =
\frac{L_1\mu }{\sqrt{2\pi }}\sqrt{t^2+\left( L_2\mu p\right)
^2}\left( \frac
m\mu \right) ^{\frac 32}\left[ t^2+\left( L_2\mu p\right) ^2+\left(
L_1\mu
k\right) ^2\right] ^{-\frac 34} \\  \\
K_{\frac 32}\left( \frac m\mu \sqrt{t^2+\left( L_2\mu p\right)
^2+\left(
L_1\mu k\right) ^2}\right)
\end{array}
\end{equation}

Again, $\left( \ref{sum2}\right) $ can be differentiated term by term
in a
straightforward although tedious calculation, and recurrence formulas
for
modified Bessel functions can be used to get
\begin{equation}
\begin{array}{c}
\left. -\frac \mu 2\frac d{dt}h\left( t,
\frac{D_B}{\mu ^2}\right) \right\rfloor _{t=0}=\left. -\frac \mu
2\frac
d{dt}\left( \sum_{p=-\infty }^\infty \sum_{k=-\infty }^\infty \
^{\prime }
\frac{L_2\mu }\pi \frac t{\sqrt{t^2+\left( L_2\mu p\right)
^2}}c_k\left(
t\right) +\frac{L_2\mu }\pi c_0\left( t\right) \right) \right\rfloor
_{t=0}
\\  \\
=-
\frac{L_1L_2}{2^3\pi ^{\frac 32}}\sum\limits_{n_1=-\infty }^\infty
\sum\limits_{n_2=-\infty }^\infty \ ^{\prime }\left( \sqrt{\left(
\frac{%
n_1L_1}2\right) ^2+\left( \frac{n_2L_2}2\right) ^2}\right) ^{-\frac
32}K_{\frac 32}\left( 2m\sqrt{\left( \frac{n_1L_1}2\right) ^2+\left(
\frac{%
n_2L_2}2\right) ^2}\right)  \\  \\
-\frac{L_1L_2m^3}{12\pi }+\left. \frac{L_1L_2\mu ^3}{2\pi
t^3}\right\rfloor
_{t=0}
\end{array}
\end{equation}

\end{document}